\documentclass[aps,prb,showpacs,twocolumn,superscriptaddress]{revtex4-2}
\usepackage{ae}
\usepackage[T1]{fontenc}
\usepackage[ansinew]{inputenc}
\usepackage{amsmath}
\usepackage{amssymb}
\usepackage{color}
\usepackage{braket}
\usepackage{subcaption}
\usepackage{appendix}
\usepackage{float}
\usepackage{color}
\usepackage{graphicx}
\usepackage[font=small, labelfont=bf,
        justification=raggedright]{caption}
\usepackage[colorlinks]{hyperref}
\usepackage{epstopdf}
\usepackage{mathrsfs}
\usepackage[normalem]{ulem}

\usepackage{cancel,xcolor}

\usepackage{soul,xcolor}
\setstcolor{red}

\usepackage{cancel}

\usepackage{soul,xcolor}
\setstcolor{red}

\makeatletter
\@ifundefined{textcolor}{}
{%
 \definecolor{BLACK}{gray}{0}
 \definecolor{WHITE}{gray}{1}
 \definecolor{RED}{rgb}{1,0,0}
 \definecolor{GREEN}{rgb}{0,1,0}
 \definecolor{BLUE}{rgb}{0,0,1}
 \definecolor{CYAN}{cmyk}{1,0,0,0}
 \definecolor{MAGENTA}{cmyk}{0,1,0,0}
 \definecolor{YELLOW}{cmyk}{0,0,1,0}
\definecolor{ORANGE}{rgb}{1,0,1}
}

\@ifundefined{definecolor}
 {\@ifundefined{definecolor}
 {\@ifundefined{definecolor}
 {\@ifundefined{definecolor}
 {\@ifundefined{definecolor}
 {\usepackage{color}}{}
}{}
}{}
}{}
}{}

\usepackage{epsfig}

\def\b0{{\bf{0}}}

\begin{document}

\title
{Crossover from Wannier-Stark localization to charge density waves for interacting spinless fermions in one dimension}




\author{N. Aucar Boidi}
\affiliation{The Abdus Salam International Centre for Theoretical Physics, Strada Costiera 11, I-34151, Trieste, Italy}
\affiliation{Centro At\'{o}mico Bariloche, Instituto Balseiro, Instituto de Nanociencia y Nanotecnolog\'{\i}a CNEA-CONICET, Bustillo 9500, 8400 Bariloche, Argentina}
\email{nairaucar@gmail.com}

\author{A. Aharony}
\affiliation{School of Physics and Astronomy, Tel Aviv University, Tel Aviv 6997801, Israel}
\email{aaharonyaa@gmail.com}

\author{O. Entin-Wohlman}
\affiliation{School of Physics and Astronomy, Tel Aviv University, Tel Aviv 6997801, Israel}
\email{orawohlman@gmail.com}

\author{K. Hallberg}
\affiliation{Centro At\'{o}mico Bariloche, Instituto Balseiro, Instituto de Nanociencia y Nanotecnolog\'{\i}a CNEA-CONICET, Bustillo 9500, 8400 Bariloche, Argentina}
\email{karenhallberg@gmail.com}

\author{C. R. Proetto}
\affiliation{Centro At\'{o}mico Bariloche, Instituto Balseiro, Instituto de Nanociencia y Nanotecnolog\'{\i}a CNEA-CONICET, Bustillo 9500, 8400 Bariloche, Argentina}
\email{karenhallberg@gmail.com}


\begin{abstract}
  We study spinless fermions on a finite chain with nearest-neighbor repulsion and in the presence of a Wannier-Stark linearly-varying electric field potential. In the absence of the interaction, the eigenstates are localized for the system's sizes larger than the localization length. We present several analytical expressions for the localization length, which is proportional to the inverse of the electric field. 
Using the density matrix renormalization group numerical technique, we observe that the ground state exhibits a decrease of the occupation on the chain sites from the `bulk', with occupation 1, to the vacuum, with occupation 0. The width of this intermediate `edge' region is also inversely proportional to the electric field, increasing linearly with the strength of the nearest-neighbor repulsion. For strong interactions, the occupations in the intermediate region exhibit a charge density wave.  We also present the local density of states for sites in the `edge' region. For the non-interacting case, the spectrum shows an increasing energy-localized structure as the field is increased, which is a consequence of the uniform energy distribution of the localized states (Wannier-Stark ladder). This structure survives for small interactions, and it smears out in the strongly interacting limit.
Experimental variations of the slope of the potential (the electric field) on cold atom chains may test these predictions.
\end{abstract}


\date{\today}

\maketitle
\section{Introduction}


One-dimensional cold atom chains have been the center of much recent experimental and theoretical interest~\cite{1,2}. In particular, such chains can be used to study the edge that separates the fully occupied insulating `bulk' and the empty vacuum. Such an edge can be generated by a confining potential, which increases as one moves away from the bulk. Recently, we have studied such confinements caused by a linear potential, i.e., a constant electric field~\cite{HF,coexistence}. 
It has been shown that the competition between linear potential and electron-electron interactions can produce a rich coexistence of locally ordered structures along the `edge'. Here we concentrate on the special case of spinless particles. 


Even without interactions, the simultaneous effects of a periodic potential and a static electric field on electrons have been an important topic in condensed matter physics~\cite{Anderson63,Taylor70,AM76}. This research started many years ago when Wannier (among many others) argued that the application of an electric field $F$ turns the continuum of Bloch states induced by the periodic potential into an equally spaced ladder of discrete energy states, with the spacing given by $eaF$~\cite{Wannier59}. Here $-e$ ($e>0$) is the electron charge, and $a$ is the lattice constant. This has  been known as the Wannier-Stark (WS)
ladder. The WS eigenstates are localized, leading to the corresponding Bloch oscillations~\cite{Anderson63,Taylor70,AM76,GKK02}. The Bloch oscillations are the counterparts in the time domain of the stationary states represented by the WS ladder in the energy domain. While for some time the existence of the WS ladder and Bloch oscillations was controversial, they are now firmly established; they have been observed in semiconductor devices~\cite{Waschke93,MB93,Guo2020,Wang2021} and accelerated optical lattices~\cite{Dahan96,Wilkinson96,AK98,Yao2020,Lang2022}.


Theoretically, the energy spectrum and eigenfunctions of {\it non-interacting} tightly bound electrons in a uniform electric field have been studied for finite and infinite chains~\cite{Fukuyama73}. In the latter case, both the WS energy spectrum and the associated eigenfunctions have been obtained analytically in terms of Bessel functions. In the present paper, we extend these analytical results for the non-interacting WS localization length, the site occupations of the fermions in the ground state of the many-particle states, and the localized density of states. We show that the analytical expressions for the infinite chain are fully recovered for sufficiently long finite chains. All the localization lengths turn out to be inversely proportional to $F$.


  While in the case analyzed in Ref.~\cite{Fukuyama73} all the eigenstates are more than exponentially localized, some recent papers claimed the existence of mobility edges in generalized WS lattices, termed ``mosaic" WS lattices ~\cite{DZ22,Gao23}. This is unexpected, as mobility edges separating localized from extended states are prevented in disorder-free low-dimensional systems~\cite{SS89}. However, the issue seems to have been settled, as it has been rigorously found that the energy spectrum is discrete even for these ``mosaic" WS lattices, except a few isolated extended states~\cite{Longhi23}.


The main features of WS physics have been recently extended from the electron to the phonon domain in Ref.~\cite{Merlin24}. The results suggested that the classical motion of a one-dimensional chain of atoms coupled through a specific force function that depends on position shows features very similar to the electronic WS counterpart, for instance localized phonon modes and phonon Bloch oscillations.


The main subject of this paper concerns the {\it interacting} Wannier-Stark fermionic lattice: electrons under the combination of a periodic potential, a constant electric field, and interactions.  The underlying physics in these systems is sometimes referred to as the ``Stark Many-Body Localization" (SMBL) ~\cite{Schulz19,NBY19,Morong21,Wei23}. The issue has received increasing recent attention due to its possible connection to the hot topic of Many-Body Localization (MBL), which concerns the interplay between disorder, localization, and interactions~\cite{Abanin19}. In the case of SMBL, the systems are disorder-free, but they seem to exhibit non-ergodic features and then constitute a class of generic nonrandom systems that fail to thermalize. Thus, the main physical feature that characterizes MBL, is that it is localization, and not disorder, that matters for attaining non-ergodic behavior. In the present case, the particle localization is provided by the electric field.


At this time, it is not clear how the MBL phenomena relate to the details of the static many-body wave function of the electrons. In particular, some of the above references find such non-ergodic phenomena only when they remove some degeneracy by adding non-linear terms to the confining potential.  In the present paper, we do not dwell on the non-ergodic features but concentrate on the characterization of the many-body wave functions. In particular, we study how the localization length induced by the electric field changes with interactions. In preparation for this question, we present an extensive analysis of the WS localization length for the non-interacting case, by analytical and numerical means.


Specifically, we study a tight-binding model of interacting spinless particles, on a finite one-dimensional lattice of $L$ sites and open boundary conditions. Setting the confining linear electric potential to vanish at the center of the chain, the resulting particle-hole symmetry between the two sides of the chain yields an average half-filling for the whole system. Without interactions, the linear potential is known to have the WS localized eigenstates. For a tunneling amplitude $t$ between nearest-neighbor sites, dimensional analysis shows that the localization length (and any other length) characterizing these states must be proportional to $t/(eaF)$.  Below we give explicit expressions for the pre-factors. The sites of the chain are fully occupied at the far left of the chain (large negative local potential), representing the `bulk', and fully empty at the far right of the chain (large positive local potential), representing the vacuum. The intermediate region, in which the site occupation decreases from 1 to 0, forms a domain wall between the fully occupied `bulk' and the completely empty vacuum. We refer to this `active' region as the 'edge'. For the non-interacting case, we show that the width of this intermediate domain is directly related to the single-particle localization length, growing linearly with $t/(eF)$.  We show that finite interactions broaden the intermediate region between the bulk and the vacuum, which still grows linearly with $t/(eF)$ with a pre-factor that depends on the tunneling and the interaction.
For strong interactions, the repulsion between nearest-neighbor sites generates a charge density wave (CDW) within the domain wall. At half-filling, the occupations along the intermediate domain are $...101010...$. 
Similar effects due to nearest-neighbor interaction were found in earlier papers~\cite{HF,coexistence,ching}.  We study the crossover between the different regimes.


Except for the analytic expressions derived for the infinite chains, all our results for the finite chains are numerical, using the density matrix renormalization group (DMRG) method \cite{white,hallberg}. This method allows us to find the ground state (GS) of the interacting many-body system for a finite one-dimensional lattice. In the non-interacting case, this GS is the Slater determinant of the single-particle WS-occupied states. We present results for different sizes of the system, all of which have occupation 1 on the left and 0 on the right so that we avoid finite size effects. Specifically, we used the matrix product state \cite{schollwoeck} implementation of the DMRG to perform the calculations, keeping around five hundred states and less than a dozen sweeps in each calculation.


Section \ref{II} defines our model. Results for the non-interacting case are given in Sec. \ref{III}, which compares analytical expressions with numerical estimates for the site occupations, different compatible definitions of the localization length, their relation to the size of the `domain walls', and the local density of states. The results for the same quantities in the interacting case are given in Sec. \ref{IV}, where the occupations in the center of the chain develop into a charge density wave.
Section \ref{conc} contains our conclusions.


\section{The Model}
\label{II} 

\begin{figure}
\includegraphics[width=0.45\textwidth]{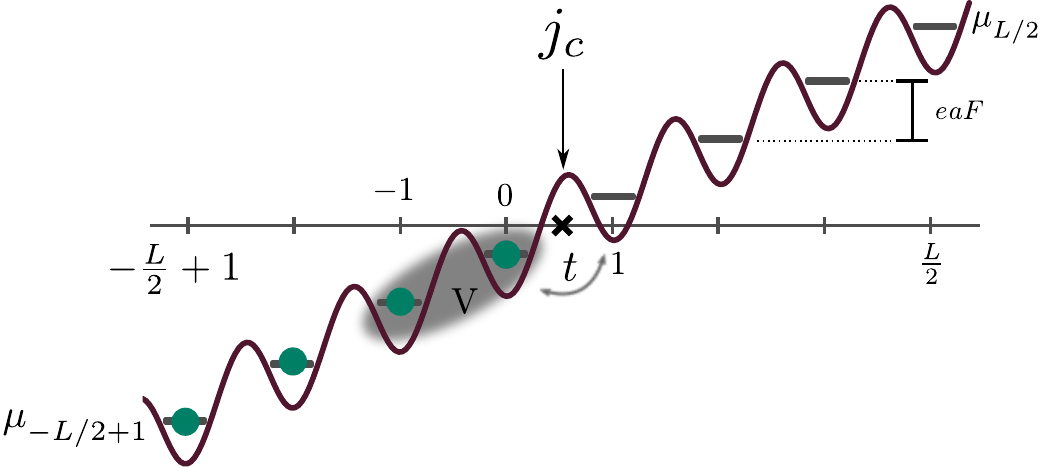}
\caption{(color online) A sketch of the one-dimensional chain with an even number of sites placed in a linear potential of slope 
$eaF$. The potential vanishes at the central link of the chain.}
\label{sketch}
\end{figure}

We study  interacting spinless fermions described by a tight-binding  Hamiltonian on an open one-dimensional lattice  of $L$ sites with lattice parameter $a$, 
\begin{align}
\hat{H}=\hat{H}_t+\hat{H}_{pot}+\hat{H}_{int}\ .
\label{int}
\end{align}
The tunneling Hamiltonian is 
\begin{align}
\hat{H}_t =&-t\sum_{j}\left( \hat{c}^{\dagger}_{j+1}\hat{c}^{}_j + 
\hat{c}^{\dagger}_j \hat{c}^{}_{j+1} \right)\ .
\label{Ht}
\end{align}
Here  $t$ is the nearest-neighbor tunneling amplitude, and $\hat{c}^{\dagger}_j$ creates a fermion at site $j$. These fermions are subjected to a linear potential, 
\begin{align}
\hat{H}_{pot}=&\sum_{j} \mu^{}_j \; \hat{c}^\dagger_j \hat{c}^{}_j  \ ,
\label{Hpot} 
\end{align}
where
\begin{align}
\mu^{}_j=  eaF\left( j - j_c \right)\ ,
\label{mu}
\end{align}
is the site-dependent potential due to the external electric field.
This potential is set to vanish at the center of the chain, $j^{}_c$, to ensure particle-hole symmetry between the two halves of the system. In the case of even (odd) $L$, the potential vanishes at the central link (site), and $j_c = 1/2$ ($j_c=0$). Our results are presented for an even $L$. The fermions interact via the repulsive nearest-neighbor interaction, $V$,
\begin{align}
\hat{H}_{int} = V\sum_{j} (\hat{n}^{}_j-1/2)(\hat{n}^{}_{j+1}-1/2)  ,
\label{Hint}
\end{align}
where $\hat{n}^{}_j=\hat{c}^\dagger_j\hat{c}^{}_j$ is the occupation operator.
 A sketch of the system is portrayed in Fig.~\ref{sketch}.

\section{The non-interacting case: Analytical Results}
\label{III}


\subsection{ The Bessel transformation}
Except for a constant shift in the  linear potential, the Hamiltonian in the absence of the repulsion reads
\begin{align}
\hat{H}_0=-t\sum_j \left( \hat{c}^{\dagger}_{j+1}\hat{c}_j^{} + \hat{c}^{\dagger}_j \hat{c}_{j+1}^{} \right) + eaF \sum_j j \hat{c}^\dagger_j \hat{c}_j^{}\ ,
\label{H0}
\end{align}
For the infinite chain, the well-known~\cite{Fukuyama73} eigenvalues of $\hat H_0$ are $E_m=eaFm$, with an integer $m$, 
and eigenfunctions 
$| m \rangle = \hat{b}^\dagger_m|{vac}\rangle$, where $|{vac}\rangle$ is the vacuum, and 
\begin{align}
    \hat{b}^\dagger_m=\sum_{j=-\infty}^\infty\mathscr{J}^{}_{j-m}[2t/(eaF)]\hat{c}^\dagger_j\ .
\label{EF}
\end{align}
Here, $\mathscr{J}_{j-m}(x)$ is the cylindrical Bessel function of the first kind of order $j-m$. Thus, $\hat H_0$ is diagonalized, 
\begin{align}
\hat{H}_0= eaF\sum_m m \; \hat{b}^\dagger_m \hat{b}_m^{}\ .
\label{Fbb}
\end{align}
In the limit of large $F$,  $\mathscr{J}_{j-m}[2t/(eaF)]\rightarrow \delta^{}_{j,m}$, and $\displaystyle \hat{b}^\dagger_m \rightarrow \hat{c}^\dagger_j$, with $m=j$.
 Note that while $m$ is an {\it energy}-related index, $j$ is a {\it site}-related index. Interestingly, both become intermixed in the building of the non-interacting eigenstates in Eq.~(\ref{EF}), and finally they merge to become the {\it same} index for strong electric fields.
 

In the following, the tunneling amplitude $t$ is set to be the energy scale, and in most of the equations, $e = a = \hbar=1$, except where they are needed for clarity. 

\subsection{The charge distribution on the chain}

\begin{figure}
\includegraphics[width=0.45\textwidth]{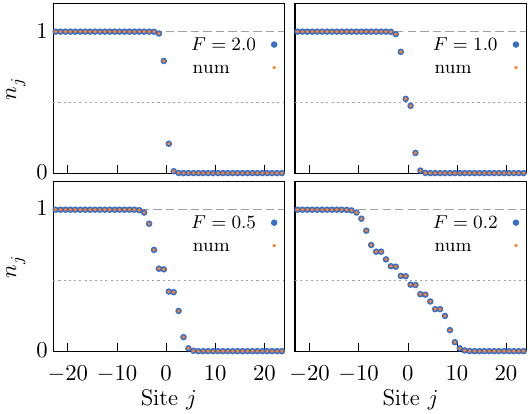}
\caption{(color online) Analytical [Eq.(\ref{nj})]
and numerical (DMRG) results for $n^{}_j(2/F$) for different $F$'s values and $L=48$. Blue (orange) dots correspond to analytical (numerical) results. The two horizontal dashed lines denote the average half-filling occupancy $\bar{n}=1/2$ and $\bar{n}=1$.}
\label{fig2}
\end{figure}

For a chain of an even number of sites $L$, and for $m=-(L/2+1),\ldots,0$ in Eq. (\ref{Fbb}), the chain is half-filled, and
the ground-state $|GS \rangle$ is 
\begin{align}
    |GS \rangle = \hat{b}_{-\frac{L}{2}+1}^{\dagger} \; \hat{b}_{-\frac{L}{2}+2}^{\dagger} \cdots \hat{b}_{-1}^{\dagger} \hat{b}_{0}^{\dagger}  |vac \rangle \ .
\label{gs}
\end{align}
In the limit $L \rightarrow \infty$, Eq.~(\ref{EF}) yields
\begin{align}
    \hat{c}^{\dagger}_j\equiv\sum_{m=-\infty}^\infty \mathscr{J}^{}_{j-m}(x)\hat{b}^\dagger_m\ ,
\end{align}
with $x=2/F\equiv 2t/(eaF)$. Then
the site occupation function of the non-interacting infinite chain  is \cite{Fukuyama73}
 \begin{align}
n^{}_j (x) =\langle GS | \; \hat{c}^\dagger_j \hat{c}^{}_j |GS\rangle= &
\sum_{m=-\infty}^0 \mathscr{J}^2_{m-j}(x)\nonumber\\
& \equiv
\sum_{m=-\infty}^0 p_j(m)
\; .
\label{nj}
\end{align}
Here, $p^{}_j(m)$ is the probability of finding the electron in the single-particle state $|m\rangle$.
Using the summation rules of Bessel
functions \cite{AS}, 
\begin{align}
\sum_{m=-\infty}^\infty \mathscr{J}^{}_{m}(x)\mathscr{J}^{}_{j+m}(x)=\delta^{}_{j,0} \ ,
\label{jjj}
\end{align}
one obtains simpler expressions for the occupations at sites closer to
the center of the chain. For instance, for the two sites around the
middle of the chain ($j^{}_c=1/2$), one finds
\begin{align}
n^{}_0(x) = [1+\mathscr{J}^2_{0}(x)]/2 \ ,\   n^{}_1(x) =
[1-\mathscr{J}^2_{0}(x)]/2 \ .
\end{align}
We show below that, interestingly,  these analytic
expressions for the non-interacting occupation numbers in the {\it
infinite} chain are excellent approximations for the numerical results
for a {\it finite} chain, provided that $L$ is larger than the size of the intermediate `edge' region. Below, we use Eq. (\ref{nj}) for $j\geq 0$, since for negative $j$, 
\begin{align} 
n^{}_{-j} (x)& = \sum_{m=-\infty}^0 \mathscr{J}^2_{m+j}(x)
 \nonumber\\
 &= \sum_{m=-\infty}^{-j} \mathscr{J}^2_{m+j}(x) + \sum_{m=-j+1}^0
\mathscr{J}^2_{m+j}(x) \ ,
\end{align}
and renaming $m^{\prime}=m+j$ yields
\begin{align}
    n^{}_{-j} (x) =&
 \sum_{m^{\prime}=-\infty}^0 \mathscr{J}^2_{m^{\prime}}(x) +se
\sum_{m^{\prime}=1}^j \mathscr{J}^2_{m^{\prime}}(x) \nonumber\\
& =
 \mathscr{J}^2_0(x) + \frac{1-\mathscr{J}^2_0(x)}{2} + \sum_{m=1}^j
\mathscr{J}^2_{m}(x) \nonumber \\
 =& n_0(x) + \sum_{m=1}^j \mathscr{J}^2_{m}(x) = 1 - n_j(x) + \mathscr{J}^2_{j}(x)\nonumber\\
 &\equiv 1-n^{}_{j+1}(x) \ .
\label{negativej}
\end{align}
This result, $n^{}_{-j}=1-n^{}_{j+1}$, reflects the anti-symmetry of the plots in Fig. \ref{fig2} with respect to the point 
$j_c=1/2$. This anti-symmetry stems from the particle-hole symmetry at half-filling and is found below also for the interacting case. In practice, this implies that it suffices to perform the calculations only for $j>0$. 
Since for $j \gg 1$, $\sum_{m=1}^j \mathscr{J}^2_{m}(x) \rightarrow
[1-\mathscr{J}^2_0(x)]/2$, it is seen that $n^{}_{-j}(x) \rightarrow 1$
for $j \gg 1$, as it should be.
The results can also be used to prove directly the average half-filling,
\begin{align}
    \sum_{j=-L/2+1}^{L/2}n^{}_j(x) \equiv L/2\ .
\end{align}


Figure~\ref{fig2} compares the site occupations derived analytically, Eq. (\ref{nj}) for $j\geq 0$, and Eq.~(\ref{negativej}) for $j<0$, with numerical results obtained by calculating the local occupation $\braket{n_j}$ for each site $j$, using the ground state, Eq. (\ref{gs}), found by the DMRG. The agreement is impressive.
An interesting feature of the plots is that for some particular values of the electric field, certain tiny plateaus seem to appear between two successive values of $j$, for instance for $F = 0.5$ in the lower left panel of Fig.~\ref{fig2}.
The reason for that becomes clear from the expression $n^{}_{j}(x) \simeq n^{}_{j-1}(x) -\mathscr{J}^2_{j-1}(x)$. If, for some $F$, the argument $x=2/F$ is close to a zero of the Bessel function of order $j-1$, $n^{}_{j-1}$ and $n^{}_j$ become arbitrarily close, and a kind of plateau appears. For $F = 0.5$, $x = 4$  is close enough to the first zero of the Bessel function, $\mathscr{J}^{}_{1}(3.8317) = 0$. Since $n^{}_{-1}(x) = n^{}_{0}(x) + \mathscr{J}^2_{1}(x)$, the same pseudo-plateau appears between sites 0 and $-1$.


An important message embodied behind the excellent agreement seen in Fig.~\ref{fig2} is that, for sufficiently large $L$, {\it no finite-size effects appear in the numerical calculations}, which are necessarily performed on finite chains ($L=48$ in Fig.~\ref{fig2}). This is because the intermediate region with non-integer filling is totally contained inside the finite system, while the inert regions with filling 1 and 0 at the extremes of the chain can be replaced by an infinite chain with no consequence on the physics in the bulk. For that reason, we also obtain an excellent fit when we truncate the sum in Eq. (\ref{nj}) to $m\geq-L/2+1$; the omitted terms are too small to matter.


\subsection{Charge distribution in the continuum limit}


Using an integral representation \cite{Martin08} for the squares of the Bessel functions, an exact formula can be derived for $n_{j}$ when the subscript $j$ in Eq. (\ref{nj}) becomes continuous,
\begin{align}
&n^{}_j\left( \frac{2}{F} \right) = \frac{1}{2} -\nonumber\\
&- \frac{1}{2\pi} \int_0^\pi \mathscr{J}_0\left[\frac{4}{F}\sin\left(\frac{\theta}{2}\right)\right]~\frac{\sin\left[\left(j-1/2\right) \theta \right]}{\sin\left(\theta /2\right)} \; d\theta \ .
\label{continuous}
\end{align}
Details of the derivation are given in Appendix \ref{eq17}. Note that $n^{}_{1/2}(2/F) = 1/2$, independent of the strength of the electric field. For $4/F \rightarrow 0$,  the zero-order Bessel function in the integrand of Eq.~(\ref{continuous}) approaches 1, and 
$n^{}_0(2/F \rightarrow 0) \rightarrow 1$, while $n^{}_1(2/F \rightarrow 0) \rightarrow 0$.
These limits imply a finite length of the intermediate region. 


\begin{figure}[h!]
\includegraphics[width=0.4\textwidth]{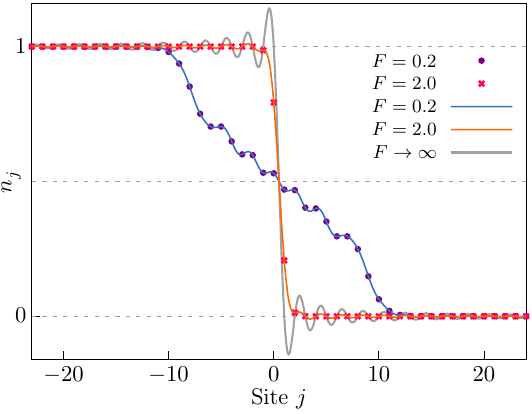}
\caption{(color online) Charge occupation $n_j(2/F)$  for discrete $j$, Eq.~(\ref{nj}) (dots) and for continuous $j$ Eq.~(\ref{continuous}) (lines), for $F=0.2$ (blue), and $F=2$ (red). The limiting curve for $F\to \infty$, Eq.~(\ref{continuous}), is also included (grey line).
}
\label{occCont}
\end{figure}

Figure \ref{occCont} compares numerical results with those derived from Eq.~(\ref{continuous}),  for $F=0.2$, $F=2$ and $F \to \infty$. The numerical numbers practically coincide with those derived from Eq.~(\ref{nj}). As seen, Eq. (\ref{continuous}) produces an oscillatory behavior that increases with $F$. These oscillations are not reflected in the occupations for integer $j$'s. 


\subsection{Localization lengths}

\label{loc}


In all panels in Fig. \ref{fig2} and Fig. \ref{occCont} there appear three regions: on the left -  the occupied `bulk' segment, with $n^{}_j=1$, on the right - the empty `vacuum', with $n^{}_j=0$, and the intermediate zone in which $n^{}_j$ decreases monotonically from $1$ to $0$. Here, we estimate the width of this intermediate region, which may be considered as a domain wall in the occupation number profile. Dimensional analysis shows that this width must be proportional to $t/(eF)$. We present several possible definitions for it and show that they are indeed all proportional to $t/F$, i.e., inversely proportional to the electric field.


\subsubsection{Estimation based on the bandwidth}
\label{QAE}

\begin{figure}
\includegraphics[width=0.4\textwidth]{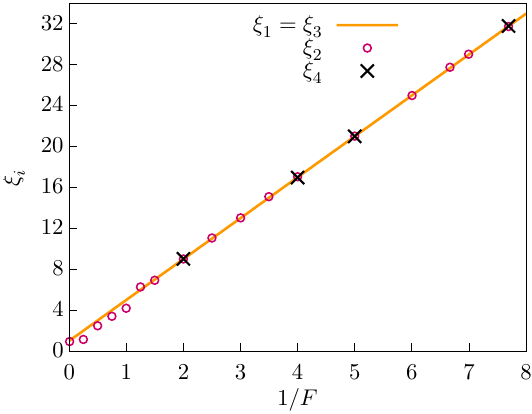}
\caption{(color online) Characteristic (localization) lengths as functions of $1/F$. 
Orange line, $\xi_1$ [Eq.~(\ref{D1})] and $\xi_3$ [Eq.~(\ref{D3})]; open circles, $\xi_2$ [Eq.~(\ref{D2})]; crosses, $\xi_4$ [Eq.~(\ref{D4})]. }
\label{charactlength}
\end{figure}

One way to estimate the width of the intermediate-filling region follows by adapting the argument of
Ref. \cite{coexistence}: 
For the non-interacting fermions, a certain site $j$ is occupied when 
$\mu^{}_j \lesssim - \; 2t$ band edge of the non-interacting system)  and is empty for
$\mu^{}_j \gtrsim 2t$. Exploiting Eq. (\ref{mu}), the two cases  can be presented as
\begin{align}
\mu^{}_< &= eaF(j^{}_<-j^{}_c) = -2t\ ,\nonumber\\
 \mu^{}_> &= eaF(j^{}_>-j^{}_c) = +2t\ ,
\label{18}
\end{align}
leading to an intermediate region of length
\begin{align}
    j^{}_> - j^{}_< 
    =\frac{4t}{eaF} \ ,
    \label{dw}
\end{align}
($4/F$ in our dimensionless unit scheme).
This estimate has been previously derived in Ref. \cite{Fukuyama73}, based on numerical analysis. 
As the derivation suggests, Eq.~(\ref{dw}) works well when the size of the domain wall involves several sites of the chain; however, it 
approaches zero in the strong electric field limit. As
the domain wall cannot be narrower than the lattice parameter (1 in our dimensionless units), we propose to define the localization length as
\begin{align}
\xi^{}_1= \frac{4}{F} + 1 \ .
\label{D1}
\end{align}
$\xi^{}_{1}$ is  displayed by the orange full line in Fig.~\ref{charactlength}.


From Eq.~(\ref{nj}) for the single-particle filling it follows that the region with intermediate filling begins at the site where there is a finite probability of finding a hole [1-$\mathscr{J}^2_{m-j}(x)$] and ends at the site with zero probability of finding an electron [$\mathscr{J}^2_{m-j}(x)$]. The distance between these two points determines the width of the single-particle wave function for a given quantum number $m$. Hence, there is a one-to-one correspondence between the width of the single-particle wave function and the width of the intermediate region $\xi^{}_1$.


\subsubsection{Estimation based on the intermediate-filling region}


An alternative estimation of the intermediate-filling region width is obtained by taking into account all sites with occupation $\varepsilon < n^{}_j < 1 - \varepsilon$, where $\varepsilon \ll 1$. Whether we use Eq.~(\ref{continuous}) for the occupation $n^{}_{j}$ at continuous $j$ or the $n^{}_j$ obtained by the DMRG, we can define $j^{}_R > 0$ as the largest site $j$ whose filling obeys
\begin{align}
n^{}_{j^{}_R} \geq \varepsilon\ .
\label{eps}
\end{align}
The corresponding size of the domain wall is then 
\begin{align}
    \xi^{}_2 = 2 \; j^{}_R \; ,
    \label{D2}
\end{align}
with the factor of 2 accounting for the chain sites with negative $j$. The results of  Eq.~(\ref{D2}) using Eq.~(\ref{continuous}) for the occupation are displayed by open circles in Fig.~\ref{charactlength}.
As seen, $\xi^{}_2$ approaches $1$ in the strong-field limit without any adjustment, as expected. Also, note that the characteristic length $4/F$ appears quite naturally in the argument of the Bessel function of order zero in Eq.~(\ref{continuous}), controlling the range where the occupation function $n^{}_j$ deviates from 1 at the left and from 0 at the right (Fig.~\ref{occCont}).


\subsubsection{Single-particle localization length}


A natural definition of the localization length of a single-particle wave function is given by its average width, for example, for a particle with the quantum number $m=0$ (the definition being independent of this quantum number), 
\begin{align}
    \tilde{\xi}=\sqrt{\sum_{j=-\infty}^\infty j^2 p_j(0)}\ ,
    \label{xiMS}
\end{align}
where $p^{}_j(0) = \mathscr J_j^2(2/F)$ is the probability of finding the single electron at site $j$ when it has been created in the Wannier-Stark eigenstate $| m=0 \rangle = \hat{b}_0^{\dagger} | vac \rangle$ [see Eq. (\ref{nj})]. This is a particular case of the more general expression,
\begin{align}
    p_j(m) = | \langle {vac} | \hat{c}_j \hat{b}_m^{\dagger} | {vac} \rangle |^2 = \mathscr J_{j-m}^2(2/F) \; , 
\end{align}
which is the probability of finding the single electron at site $j$, when created in the eigenstate $| m \rangle = \hat{b}_m^{\dagger} | vac\rangle$. 
Using the Bessel functions'  identity \cite{AS}
\begin{equation}
\mathscr J_{j-1}^{}(x)+\mathscr J_{j+1}^{}(x)=(2j/x)\mathscr J_{j}^{}(x)
\end{equation}
 and Eq. (\ref{jjj}) derived for the infinite chain, 
we obtain 
\begin{align}
\tilde{\xi}^{2}(x) = \sum_{j=-\infty}^{\infty} |j \mathscr J_{j}^{}(x)|^2 = 2(x/2)^2 = 2(t/F)^2\ .
 \label{xtilde}
\end{align}
$\tilde{\xi}$ is inversely propotional to the electric field, $t/F$, as expected from dimensional analysis. 
Interestingly enough, defining   
\begin{align}
 \xi^{}_3(x) = 2\sqrt{2} \; \tilde{\xi}(x) + 1\; ,
\label{D3}
\end{align}
the single-particle localization length as defined in Eq. (\ref{xtilde}) becomes identical to $\xi^{}_{1}(x)$, Eq. (\ref{D1}). Our results show that $\xi_3$, which represents the well-documented exponential localization of the Wannier-Stark eigenstates, together with $\xi_1$ and $\xi_2$, which represent the width of the domain wall of the occupation number function, are just different ways of quantifying the {\it same} Wannier-Stark localization physics.  See,  however, Ref. \cite{Note}.


\subsubsection{Many-particle localization length}


We define the many-particle localization length as the mean-square distance of the charge distribution:
\begin{align}
\tilde{\tilde{\xi}}=2\sqrt{\sum^{}_{j>0}  (j-j^{}_c)^2\tilde{p}_j}\  ,
\label{xinum}
\end{align}
where $p^{}_j(0)$ in Eq. (\ref{xiMS}) is replaced by $\tilde{p}^{}_j\equiv n^{}_j/{\cal N}$, with ${\cal N}=\sum_{j'>0}n^{}_{j'}$ being the total probability of finding electrons on the right-hand side of the chain (equal to the total number of holes on the left-hand side).  Note that $j^{}_c=1/2$ for even $L$. 
The sum runs only on $j>0$, where $n^{}_j$ decreases from $1/2$ to zero, the factor of 2 accounting for the negative $j$'s.
We shall use the same definition when including the interaction between the fermions in the discussion.


The analytic expression for $\tilde{\tilde{\xi}}$ for non-interacting fermions is worked out in Appendix \ref{AAA}. It is compared with the DMRG results for the ground state of non-interacting fermions ($V=0$) in Fig. \ref{charactlength}, utilizing for convenience the expression
\begin{align}
\xi^{}_4 = A_4 \tilde{\tilde{\xi}}+ B_4 \; ,
\label{D4}
\end{align}
 where the constants $A_4=2.13$ and $B_4=0.88$ are adjusted for fitting it with
Eq.~(\ref{D1}). The matching of the field-dependence of $\xi^{}_4$ with the other lengths justifies using Eq. (\ref{D4}) also for estimating the localization length of interacting fermions, carried out in Sec.~\ref{IV}.


\subsection{The local density of states for non-interacting fermions}

\begin{figure}
\begin{subfigure}{0.4\textwidth}
\includegraphics[width=1\textwidth]{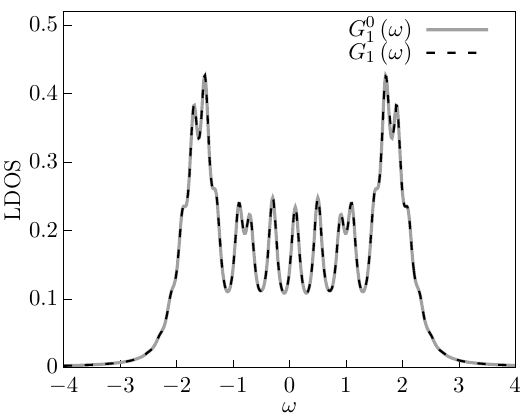}
\end{subfigure}
\begin{subfigure}{0.4\textwidth}
    \includegraphics[width=1\textwidth]{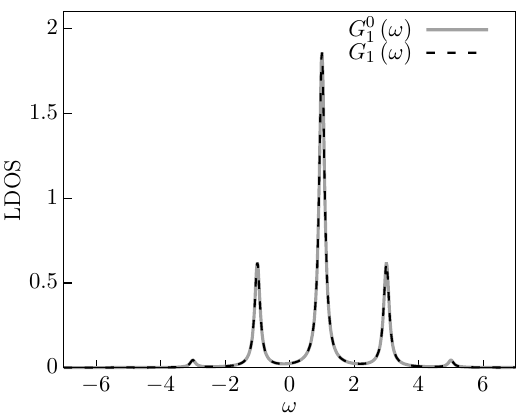}
\end{subfigure}
\caption{Local density of states (LDOS): The analytic expression for  $G^{0}_1(\omega)$ for $j=1$ [full line, Eq.~(\ref{rhoj1})] and the numerical result for $G_1(\omega)$ [dashed line] for a chain with $L=48$ and $V=0$ ($\eta=0.1$). Upper panel: $F=0.2$; lower panelt: $F=2$.}
\label{LDOSF2F0.2}
\end{figure}

The local density of states (LDOS) is calculated by
applying Zubarev's equation-of-motion method \cite{Taylor70} to the Hamiltonian $\hat{H}^{}_0$, Eq.~(\ref{H0}). The frequency-dependent diagonal Green's function gives the site-dependent density of states
\begin{align}
    G^{0}_j(\omega) = -\frac{1}{\pi} \; {\rm Im} \langle \langle  \hat{c}^{}_j;\hat{c}_j^{\dagger} \rangle \rangle \; .
\label{rhojIm}
\end{align}
This Green's function obeys the Dyson equation \cite{Taylor70}
\begin{align}
    \omega \; \langle \langle  \hat{c}^{}_j;\hat{c}_{j'}^{\dagger} \rangle \rangle& = 
    [\hat{c}^{}_j, \hat{c}_{j'}^{\dagger}]_+   +
    \langle \langle [\hat{c}^{}_j,\hat{H}_0]_\pm; \hat{c}_{j'}^{\dagger}   \rangle \rangle\nonumber\\
    & = 
    \delta^{}_{j,j'} + \langle \langle [\hat{c}^{}_j,\hat{H}^{}_0]_\pm; \hat{c}_{j'}^{\dagger}   \rangle \rangle \; .
\label{eqmot}
\end{align}
Here, $[\hat{A};\hat{B}]_{\pm}$ denotes  either the anti-conmutator ($+$) or commutator ($-$) between operators $\hat{A}$ and $\hat{B}$. Using the diagonal form of 
$\hat{H}^{}_0$, Eq.~(\ref{Fbb}), the Dyson equation Eq.~(\ref{eqmot}) becomes
\begin{align}
    \omega \; \langle \langle  \hat{c}^{}_j;\hat{c}_{j'}^{\dagger} \rangle \rangle&= \delta_{j,j'} + 
    F \sum_{m=-\infty}^{\infty} m \mathscr{J}_{j-m}(2/F) \nonumber\\
    &\times\sum_{j''=-\infty}^{\infty} \mathscr{J}_{j''-m}(2/F) \langle \langle  \hat{c}^{}_{j''};\hat{c}_{j'}^{\dagger} \rangle \rangle \; ,
\label{eqmot2}
\end{align}
Multiplying both sides of Eq. (\ref{eqmot2}) by $\mathscr{J}_{j'-m}(2/F)$ and summing over $j'$ yields:
\begin{align}
    \sum_{j'=-\infty}^{\infty} \mathscr{J}_{j'-m}(2/F) \langle \langle  \hat{c}^{}_{j'};\hat{c}_{j}^{\dagger} \rangle \rangle =
    \frac{\mathscr{J}_{j-m}(2/F)}{(\omega-mF)} \; .
\label{lcenter}
\end{align}
Hence, for $j=j'$ and using Eq.~(\ref{lcenter}) , Eq. (\ref{eqmot2}) yields 
\begin{align}
\langle \langle  \hat{c}^{}_j;\hat{c}_j^{\dagger} \rangle \rangle =
\sum_{m=-\infty}^{\infty} \frac{\mathscr{J}_{j-m}^{\;2}(2/F)}{(\omega-mF)} \; .
\end{align}
Replacing $\omega \rightarrow \omega + i \eta$ with $\eta \to 0$, that broadens the delta-function spikes, Eq.~(\ref{rhojIm}) becomes
\begin{align}
G^{0}_j(\omega) = \frac{\eta}{\pi} \sum_m 
\frac{\mathscr{J}^{\;2}_{j-m}(2/F)}{[\omega -(m-j_c)F]^2+\eta^2} \; ,
\label{rhoj1}
\end{align}
which coincides with the result in Ref. \cite{DW88} for $\eta \rightarrow 0$. 
For small $\eta$, this is a sum of Lorentzian functions centered at the different energy levels $(m-j_c)F$ weighted by the square of the corresponding Bessel function (`the oscillator strength'). Calculating numerically the LDOS
using the correction vector method \cite{ramasesha} to obtain the Green's function $G^{}_j(\omega)$ (details in Ref. \cite{SMWS}), results in an excellent agreement with Eq. (\ref{rhoj1}), as depicted in  Fig.~\ref{LDOSF2F0.2}.
As the electric field $F$ increases, the oscillator strength becomes concentrated on a smaller number of peaks reflecting, in the frequency domain, the associated decreasing localization in real space of the Wannier-Stark eigenstates. For large electric fields, $\mathscr{J}^{\;2}_{j-m}(2/F) \rightarrow \delta_{j,m}$ and the LDOS becomes a single-peaked (Lorentzian-like) structure, centered at frequency $\omega = (j-j_c)F$.   
The localization length is proportional to the width of the region in real space whose site energy is contained within the energy support $\Delta \omega$ of the LDOS $\sim  \Delta \omega / F$.


Finally, we mention that
\begin{align}
    n^{}_j = \int_{-\infty}^0 G^{0}_j(\omega) \; d \omega \; ,
\end{align}
is equal to the expression of the occupation number function given in Eq.~({\ref{nj}}).


\section{Interacting Fermions}
\label{IV}


Currently, no analytical results are available for the Hamiltonian (\ref{int}), which includes the interaction $V$ between the fermions. We thus resort to
numerical calculations based on the DMRG.


\subsection{Charge distribution: Crossover to a charge density wave (CDW) region}

\begin{figure}
\includegraphics[width=0.41\textwidth]{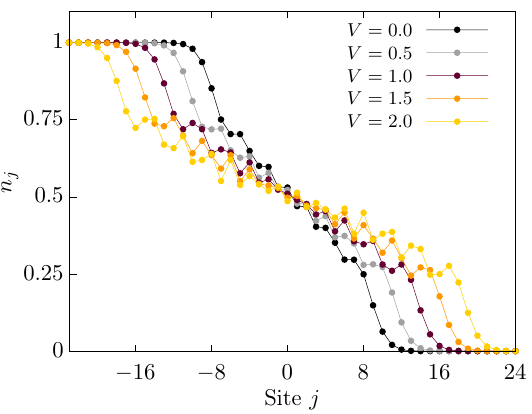}
\caption{(color online) Site occupation for an $L=48$ chain, an electric field $F=0.2$, and weak interactions.}
\label{locV}
\end{figure}
\begin{figure}
    \includegraphics[width=0.43\textwidth]{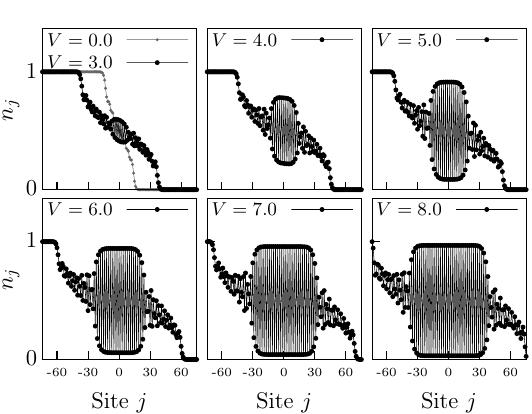}
\caption{Site occupation  for an $L=150$ chain, electric field $F=0.13$ and interactions $V\geq 2$ ($V=0$ is included as a reference).}
\label{njvsVF013}
\end{figure}
\begin{figure}
     \includegraphics[width=0.45\textwidth]{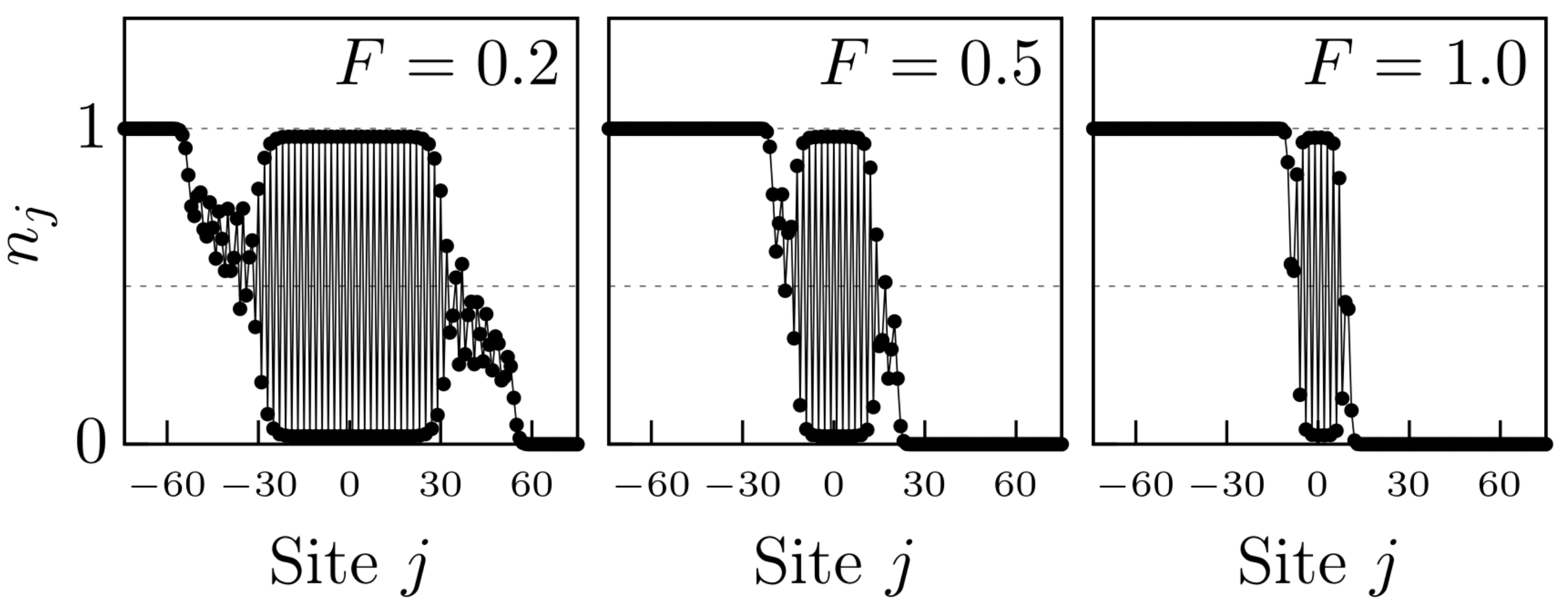}
\caption{Site occupation for an  $L=150$ chain, interaction  $V=9$,  and different values of the electric field.}
\label{njV9}
\end{figure}

The Wannier-Stark localized states characterizing the Hamiltonian in the absence of the interaction ($V=0$) progressively turn into 
a charge density wave, with occupations alternating between 1 and 0, so that $n^{}_jn^{}_{j+1}=0$, i.e., $...10101010...$ when $V$ is very strong. As might be expected, the competition between the interaction $V$ and the electric field $F$ is mainly significant around the center of the chain, where the site energy is not
too negative (at the left) or too positive (at the right). We examine this competition in Figs. \ref{locV}, \ref{njvsVF013}, and \ref{njV9}. Note that in all these plots the particle-hole symmetry relation,  $n^{}_{-j}=1-n^{}_{j+1}$, is obeyed.


Figure \ref{locV} presents the local (charge) occupations for a weak electric field $F$ and weak interactions, $V\leq 2$.   As seen, there appear small oscillations in $n^{}_j$ around the center of the chain as $V$ increases.


In the absence of an electric field, i.e., for a uniform chain, a charge density wave appears at a critical value, $V=2$, of the interaction~\cite{Gotta}. In the present case, 
the onset of the charge density wave appearing at larger values of the interaction is depicted in
Figs. \ref{njvsVF013} and \ref{njV9}, both portraying a new region of charge density wave, whose width increases with $V$.  As mentioned, three different regions can be identified in the ground state: a band-insulating phase at the left and right regions (completely full and empty sites, respectively) and an `edge' with strong CDW characteristics in the middle of the chain in between. This new region is separated from the band-insulating phases by `domain walls' with non-integer fillings. As $V$ increases, it approaches the `pure' CDW structure where $n^{}_{j}n^{}_{j+1}=0$ (see Fig. \ref{njvsVF013}). In contrast,  increasing the electric field $F$  narrows down the central CDW domain while maintaining its intensity, as depicted in Fig.~\ref{njV9}.  This behavior provides the possibility of experimentally manipulating and controlling CDW-ordered regions within a confined system.


Contrary to the results obtained in the extended Hubbard model~\cite{coexistence}, we do not observe other CDW's with different mean densities at other regions of the chain. This is a consequence of the fact that, for the uniform spinless model with nearest-neighbor interactions considered here, the only possible CDW is the one with the mean filling of 0.5~\cite{Gotta}.  Instead, for the extended Hubbard model, there are CDW's
with different periodicities at mean fillings away from 1/2, e.g., 1/4.

\begin{figure}[h!]
     \includegraphics[width=0.45\textwidth]{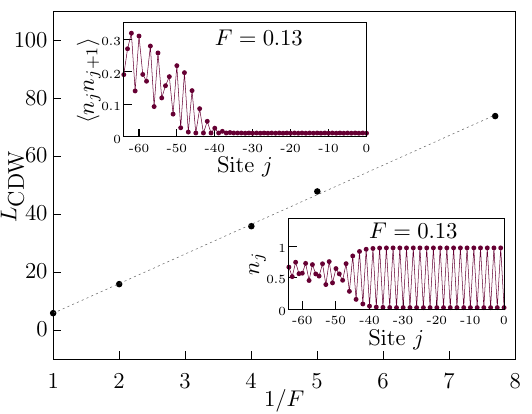}
\caption{Extension of the CDW order around the center of an $L=150$ chain as a function of $1/F$ for a large value of the interaction,  $V=9$ (main panel, the dashed line is a guide to the eye).
The top left and bottom right insets show the charge-charge correlator $\braket{n_j n_{j+1}}$ and the local occupation $\braket{n_j}$, respectively, for the region close to the transition.}
\label{Lcdw}
\end{figure}

We have calculated the size of the CDW region,  $L_{\text{CDW}}$,  and found that it has a linear dependence on  $1/F$, as shown in Fig.~\ref{Lcdw}. 
This size is obtained by calculating the charge-charge correlator $\braket{n^{}_j n^{}_{j+1}}$ (top left inset in Fig.~\ref{Lcdw}), when the order sets in this correlator is nearly constant and goes to zero for large $V$. The boundaries of this region are obtained by setting a bound $\epsilon = 0.001$ to the difference between the charge correlator and its mean value in the CDW region.


\subsection{Characteristic lengths}
\subsubsection{Derivation based on the bandwidth of the Hamiltonian with interactions}


Dimensional analysis implies that any typical length should be a linear combination of $t/eaF$ and $V/eaF$. Indeed, Wei {\it et al.}~\cite{Wei23} use arguments similar to ours in  Sec. \ref{QAE} to obtain an estimation of the width of this central region. For $V \neq 0$, the `band boundaries' can be estimated by considering the energy of having only one particle in the system with respect to the empty system, $-2t-V$, for the lower limit, and the energy of having a completely filled band with respect to having one hole in the system, $2t+V$ for the upper limit. Then, in the same spirit of Sec. \ref{QAE}, we will have an intermediate occupation when $ -2t-V < \mu_j < 2t+V $ and the interacting version of Eq.~(\ref{D1}) becomes:
\begin{align}
\xi^{}_{1,int} = 2\left( 2t + V \right)/(eF) +  c\ ,
\label{D1int}
\end{align}
where $c$ is a constant.


\subsubsection{Comparison with the intermediate-filling region $\xi_2$ [Eq.~(\ref{D2})]  }

\begin{figure}
\includegraphics[width=0.45\textwidth]{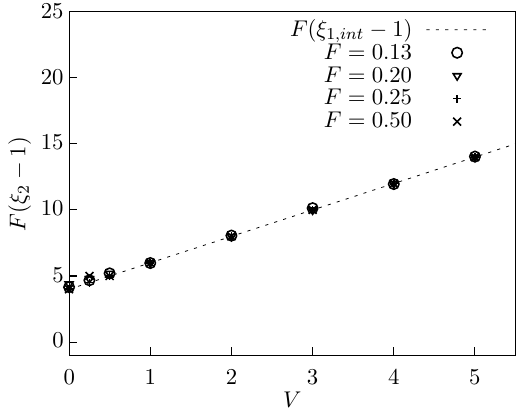}
\caption{Scaled characteristic length $\xi_2$ [Eq.~(\ref{D2})] obtained from the width of the intermediate region for $n_j$ (for $\epsilon=0.02$), for several values of field $F$ and interactions $V$ . The dashed line indicates the scaled expression for the analytical expression $\xi_{1, int}$, Eq.~(\ref{D1int}).}
\label{newxi}
\end{figure}

Figure \ref{newxi} exhibits the results of the width of the interacting intermediate region where the numerically calculated density is $0<n^{}_j<1$ and compares it with the analytical estimation described above. The scaled results for several values of the electric field $F$ and the interaction $V$ are in excellent agreement.
This agreement induces us to define a localization length also for the many-body wave function. We find two scenarios for this model: one for small interactions, where no order arises in the intermediate region, and one for larger interactions, where a CDW is formed in this region, surrounded by two domain walls to the completely empty and full regions of the chain. In both cases, we can define a `localization' length that is compatible with the one defined in the non-interacting case. This is possible because -- for strong repulsion -- the fermions are localized in the vicinity of the full and empty regions.


\subsubsection{Comparison with the many-particle localization length $\xi_4$ [Eq.~(\ref{D4})] for small $V$.}

\begin{figure}
     \includegraphics[width=0.45\textwidth]{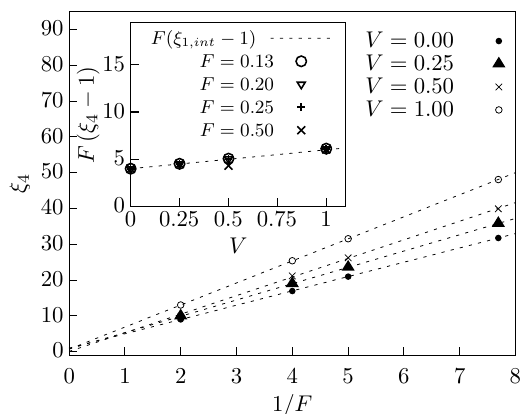}
\caption{Characteristic length $\xi_4$ [Eq.~(\ref{D4})] vs. $1/F$ obtained from the second moment of the intermediate region for $n_j$,  for several values of field $F$ and interactions $V$. The dashed lines are guides to the eye, showing a linear behavior with $1/F$.
Inset: Scaled characteristic length $\xi_4$ [Eq.~(\ref{D4})]. The dashed line in the inset indicates the scaled expression for the analytical expression $\xi_{1, int}$, [Eq.~(\ref{D1int})] exhibiting an excellent agreement.}
\label{LintV}
\end{figure}

For $V\leq 2t$ the site occupations in Fig. \ref{locV} are still qualitatively similar to those in the non-interacting case, except for the existence of density fluctuations within the intermediate region. As for this parameter range, there is no CDW, its width can be estimated using Eq.~(\ref{D4}) numerically.  Figure \ref{LintV} portrays $\xi^{}_4$ versus $1/F$  for $L=150$ and for relatively small interactions, $V=0$, 0.25, 0.5,  and 1, showing a linear behavior. Again we find an excellent agreement with $\xi^{}_{1,int}$ as seen in the scaled inset.
 

\subsubsection{Local densities of state}

\begin{figure}
\includegraphics[width=0.45\textwidth]{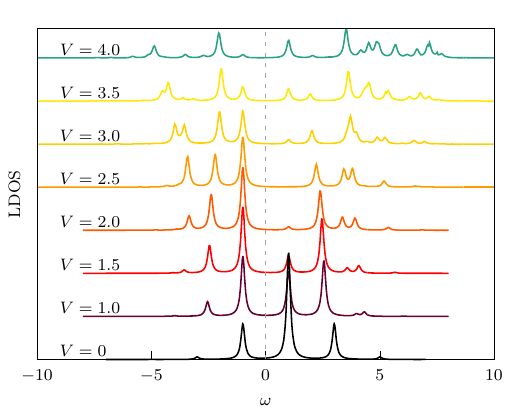}
\caption{(color online) Local densities of state $G_1(\omega)$ for several values of $V$, $L=24$, $F=2$, and $\eta =0.1$. The curves are shifted for visualization purposes, all on the same scale.}
\label{LDOSF2Vs}
\end{figure}

The effect of a weak interaction on the localization length can also be observed in the structure of the local density of states.  Figure~\ref{LDOSF2Vs} presents the evolution of the LDOS for $F=2$, $L=24$, and several values of $V$ obtained using DMRG. For the non-interacting case, as detailed before (Fig.~\ref{LDOSF2F0.2}), the LDOS has a well-defined discrete peak structure for this large value of $F$, where the peaks are separated by the WS level splitting $F$. Similar features have been observed in Ref. \cite{Udono23}.
For small $V$, this finite peak structure is modified and the total width is slightly increased, thus implying a larger localization length.


\section{Conclusions}
\label{conc}


We studied the ground state (GS) properties and local density of states (LDOS) of a spinless fermionic chain with nearest-neighbor repulsion $V$ in a linear Wannier-Stark (WS) potential. The competition between the (tunneling) kinetic energy, the repulsion, and the WS potential leads to the formation of regions with different characteristics. At the extreme left of the chain, there is a fully occupied zone, while at the extreme right, the sites are empty. On the left side, the site energies are very negative, so the sites are occupied irrespective of $V$. Similarly, on the right-hand side 
the site energies are high, preventing their occupation. 

In the middle region, there exists competition, leading to intermediate fillings with localized fermions. 
For $V=0$ we present analytical results for the GS and LDOS, which match the numerical results using the DMRG when the localization length is smaller than the system size (and no finite-size dependence is found). This agreement serves as a test for the accuracy of our numerical DMRG (approximate) calculations. 
For finite values of $V$, the intermediate region grows with $V$. When $V\leq 2t$ this region presents small filling fluctuations, while for $V>2t$, a charge-density wave (CDW) is formed. 
This happens since around the middle of the chain, the chain is close to be half-filled because the site-varying potential lies within the charge gap. In this region, a CDW order is energetically preferable when $V$ is large enough (the critical value being $V_c=2t$ for the uniform case). It is worth noting that we do not find CDWs at other average fillings. Such fillings were found in similar studies for the extended Hubbard model \cite{HF,coexistence}. This is because for the Hamiltonian studied here, a CDW forms only at half-filling for the uniform case. The size of the CDW region increases with $V$ and decreases with $F$, providing the possibility of experimentally manipulating and controlling CDW-ordered regions within a confined system.

The width of the intermediate region, between the fully occupied and the fully empty regions, is proportional to $t/F$. This is true in all the cases studied, including the localization length in the non-interacting WS ladder. In that case, the width is excellently approximated by its analytic value for the WS localization length in infinite chains. We presented several novel ways to measure this width. For the interacting case, this width is linear both in $V$, and in $1/F$. The behavior of the LDOS also reflects the localization features observed from the static properties. Experimental variations of the slope of the potential (the electric field) on cold atom chains may be able of test these predictions.


\onecolumngrid


\appendix


\section{Derivation of Eq. (\ref{continuous}) for the site occupation of a continuous $j$}
\label{eq17}


Our starting point is the integral representation of $\mathscr{J}^{2}_{n}(x)$ \cite{Martin08},  
\begin{align}
\mathscr{J}^{2}_{m-j}(x) = \frac{1}{2\pi} \int_{-\pi}^{\pi} \mathscr{J}^{}_{0}
(x,\theta) \; e^{i(m-j)\theta} d\theta\ , \ \ \ \mathscr{J}^{}_{0}(x,\theta) = \mathscr{J}^{}_{0} [2x \sin (\theta/2)]\ .
\end{align}
It follows that
\begin{align}
\sum_{m=1}^N \mathscr{J}^{2}_{m-j}(x) &= \frac{1}{2\pi} \int_{-\pi}^{\pi}
\mathscr{J}^{}_{0}(x,\theta) \sum_{m=1}^N e^{im\theta} e^{-ij\theta} d\theta= \frac{1}{2\pi} \int_{-\pi}^{\pi}
\mathscr{J}^{}_{0}(x,\theta) \frac{e^{i(N+1/2)\theta}-e^{i\theta/2}}{2i\sin(\theta/2)} e^{-ij\theta} d\theta\nonumber\\
&= \frac{1}{2\pi} \int_0^{\pi}
\mathscr{J}^{}_{0}(x,\theta) \left\{\frac{\sin[(j-1/2)\theta]}{\sin(\theta/2)} + \frac{\sin[(N+1/2-j)\theta]}{\sin(\theta/2)}\right\} d\theta \ .   
\label{A3}   
\end{align}
Since the site occupation is given by
\begin{align}
 n_j(x) = \sum_{m=-N}^0 \mathscr{J}^{2}_{m-j}(x) = 1 - \sum_{m=1}^N \mathscr{J}^{2}_{m-j}(x) \ ,
\end{align}
we find
\begin{align}
n_j(x) = 1 - \frac{1}{2\pi} \int_0^{\pi}
\mathscr{J}^{}_{0}(x,\theta) \left\{\frac{\sin[(j-1/2)\theta]}{\sin(\theta/2)} + 
\frac{\sin[(N+1/2-j)\theta]}{\sin(\theta/2)}\right\} d\theta \ .
 \label{A7}   
\end{align}
As shown in Ref. \cite{Martin08}, the second term in the curly brackets vanishes in the $N\rightarrow\infty$ limit, leading to Eq. (\ref{continuous}) in the main text. Figure \ref{fig2} indeed confirms that 
for integer $j$'s, $n_j(x)$ as given by Eq. (\ref{nj}) coincides with the one obtained from Eq. (\ref{continuous}).
Thus, for noninteger values of $j$, Eq. (\ref{continuous}) provides an interpolation scheme for the site occupations.


Interesting limits of Eq.(\ref{continuous}), based on the derivation in Ref. \cite{Martin08},  are:
a) $n_{j=1/2}(x)=1/2$. The non-interacting occupation function $n_j(x)$ is symmetric around the central point of the chain;
b) $n_{j\rightarrow \infty} \rightarrow 0$; and c) $n_{j\rightarrow -\infty} \rightarrow 1$. 


\section{Analytic expression for Eq. (\ref{xinum})}
\label{AAA}

\begin{figure}
\includegraphics[width=0.6\textwidth]{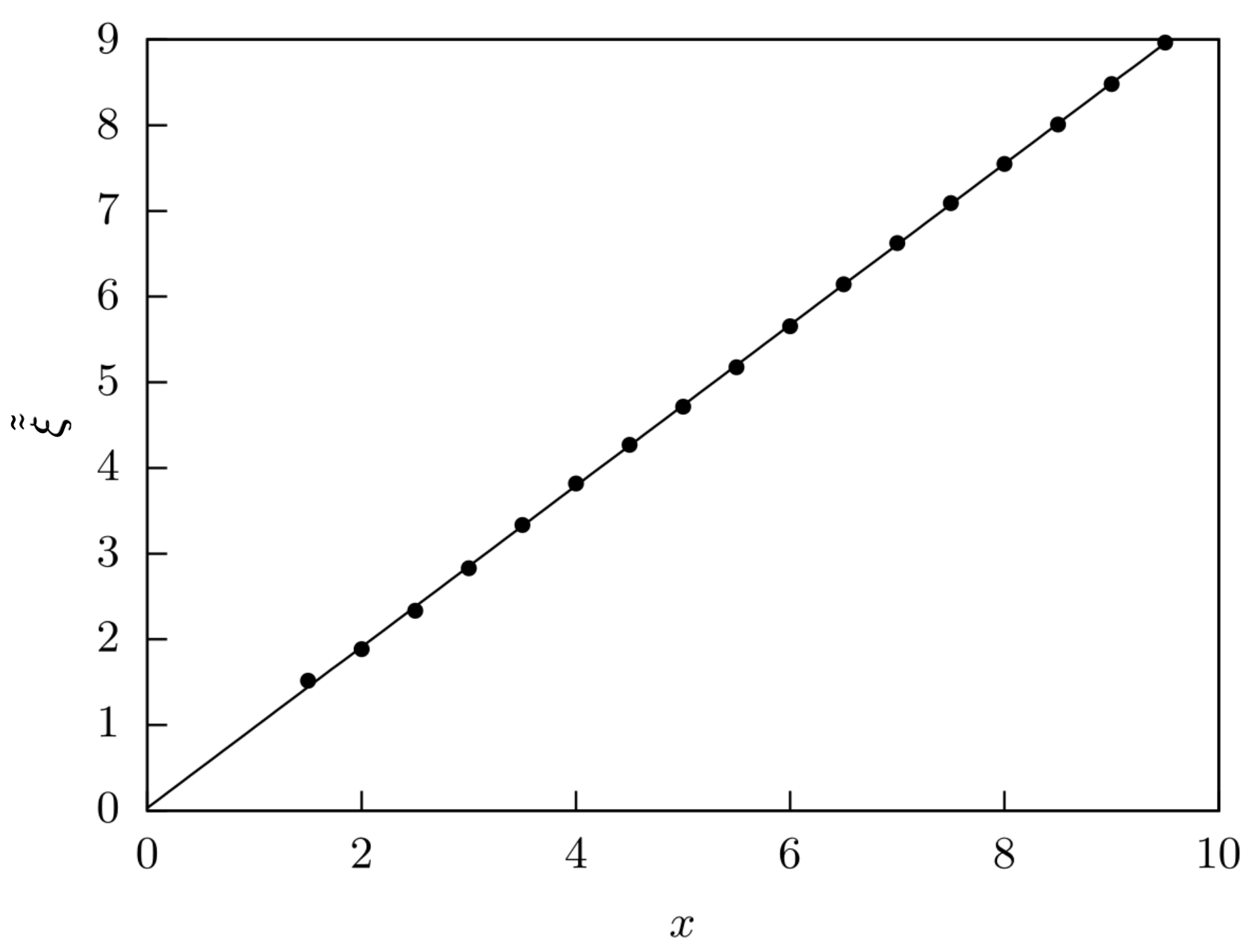}
\caption{$\tilde{\tilde{\xi}}$ as a function of $x=2/F$ from Eq. (\ref{newAAA}). The straight line is $\tilde{\tilde{\xi}}=0.9395 x$.
}
\label{newAA}
\end{figure}

We begin by rewriting Eq. (\ref{xinum}) for $j_{c}=1/2$ using Eq. (\ref{continuous}),  in the form
\begin{align}
\tilde{\tilde{\xi}}^{2}_{}=&\frac{4}{\cal N}\sum_{j>0}(j-\frac{1}{2})^{2}\Big [ \frac{1}{2} - \frac{1}{2\pi} \int_0^\pi \mathscr{J}_0\left[\frac{4}{F}\sin\left(\frac{\theta}{2}\right)\right]~\frac{\sin\left[\left(j-1/2\right) \theta \right]}{\sin\left(\theta /2\right)} \; d\theta \Big]\equiv 4(W^{}_1+W^{}_2)/\cal{N}\ , 
\end{align}
where 
\begin{align}
W^{}_1&=\frac{1}{2}\sum_{j=1}^N(j-1/2)^2=\frac{N(4N^2-1)}{24}\ ,\nonumber\\
W^{}_2&=-\frac{1}{2\pi}\int_0^\pi {\cal J}^{}_0[2 x \sin(\theta/2)]\frac{{\cal S}^{}_N}{\sin(\theta/2)}d\theta\ ,\nonumber\\
{\cal N}&=\sum^{}_{j>0}n^{}_j=\frac{N}{2}-\frac{1}{2\pi}\int_0^\pi {\cal J}^{}_0[2 x \sin(\theta/2)]\frac{\bar{\cal S}^{}_N}{\sin(\theta/2)}d\theta\ .
\end{align}
For $N=L/2\rightarrow\infty$ and
\begin{align}
&\bar{\cal S}^{}_N=\sum_{j=1}^N\sin[\theta(j-1/2)]=\frac{N}{2}-\frac{\sin^2(N \theta/2)}{\sin(\theta/2)}\ ,\nonumber\\
&{\cal S}^{}_N=\sum_{j=1}^N(j-1/2)^2\sin[\theta(j-1/2)]=
-\Big[\frac{\partial}{\partial\theta}\Big]^2\bar{\cal S}^{}_N\nonumber\\
&=\frac{-3+(3-4N^2)\cos(N \theta)+\cos(\theta)[-1+(1+4N^2)\cos(N \theta)]+4N \sin(\theta)\sin(N\theta)}
{16\sin^3(\theta/2)}\nonumber\\
&=\sin(\theta/2)\Big[-N^2\frac{\cos(N\theta)}{2\sin^2(\theta/2)}+N\frac{\sin(N\theta)\cos(\theta/2)}{2\sin^3(\theta/2)}
-\frac{[2-\sin^2(\theta/2)]\sin^2(N\theta/2)}{4\sin^4(\theta/2)}\Big].
\end{align}
Explicitly, 
\begin{align}
\frac{1}{2\pi}\int_0^\pi \frac{\bar{\cal S}^{}_N}{\sin(\theta/2)}d\theta=\frac{N}{2}\ ,\ \ \ 
W^{}_1=\frac{1}{2\pi}\int_0^\pi \frac{{\cal S}^{}_N}{\sin(\theta/2)}d\theta,
\end{align}
and thus
\begin{align}
W_1+W_2&=\lim^{}_{N\rightarrow\infty}\frac{1}{2\pi}\int_0^\pi\big(1-{\cal J}^{}_0[2 x \sin(\theta/2)]\big)\frac{{\cal S}^{}_N}{\sin(\theta/2)}d\theta\ ,\nonumber\\
{\cal N}&=\lim^{}_{N\rightarrow\infty}\frac{1}{2\pi}\int_0^\pi \big(1-{\cal J}^{}_0[2 x \sin(\theta/2)]\big)\frac{\bar{\cal S}^{}_N}{\sin(\theta/2)}d\theta\ .
\label{xixi}
\end{align}
Figure \ref{newAA} shows the numerical evaluation of 
\begin{align}
\tilde{\tilde{\xi}}=2\sqrt{\left(W_1+W_2\right)/{\cal N}}.
\label{newAAA}
\end{align}
The plot is excellently approximated by a straight line. To check the deviation from linearity, we use the expansion
\begin{align}
1-{\cal J}^{}_0[2 y)]=-\sum_{k=1}^\infty (-1)^ky^{2k}/(k!)^2.
\label{expJ}
\end{align}
For large $N$, the integrals of each term become independent of $N$, and we find
\begin{align}
\tilde{\tilde{\xi}}^2&=\frac{1}{16}x^2+\frac{3}{128}x^4-\frac{5}{4608}x^6+\frac{7}{147456}x^8-\frac{1}{655360}x^{10}+\cdots\ ,\nonumber\\
{\cal N}&=\frac{1}{4}x^2-\frac{1}{32}x^4+\frac{1}{324}x^6-\frac{5}{36864}x^8+
\frac{7}{1474560}x^{10}+\cdots.
\label{ser}
\end{align}
This does not give a linear $\xi^{}_A$, but as seen from the figure, the full expression is quite close to a straight line.


\section{Strong electric field limit}


The Hamiltonian Eq. (\ref{int}) can be expressed in terms of the operators 
$\{\hat{b}_m,\hat{b}_m^{\dagger}\}$ that diagonalize the Hamiltonian prevailing in the absence of the interaction, Eq. (\ref{H0}). Using the transformation $\{\hat{c}^{}_j,\hat{c}^{\dagger}_j \} = 
\sum_m\mathscr{J}^{}_{j-m}(x) \{\hat{b}^{}_m,\hat{b}^{\dagger}_m \} $ with $x=2/F$, one obtains
\begin{align}
    \hat{H} = F \sum_m m \; \hat{b}_m^{\dagger} \hat{b}^{}_m +
    V \sum_{m_1,m_2,m_3,m_4} f(m_1,m_2,m_3,m_4) \; \hat{b}^{\dagger}_{m_1} \hat{b}^{}_{m_2}
    \hat{b}^{\dagger}_{1+m_3} \hat{b}^{}_{1+m_4} \ ,
    \label{HA}
\end{align}
where 
\begin{align}
    f(m_1,m_2,m_3,m_4) = \sum_j \mathscr{J}^{}_{j-m_1}(x) \mathscr{J}^{}_{j-m_2}(x) \mathscr{J}^{}_{j-m_3}(x) \mathscr{J}^{}_{j-m_4}(x) \; .    
\end{align}
This Hamiltonian is simplified in the strong-field limit, i.e., for
 $x \rightarrow 0$. Then, exploiting 
the series expansion of the Bessel functions for small arguments 
\begin{align}
    \mathscr{J}^{}_{\nu}(z \rightarrow 0) \rightarrow \left( \frac{z}{2} \right)^{\nu} 
    \left[1 + \frac{(-z^2/4)}{\Gamma(\nu+2)} + ... \right] \ ,
\end{align}
the leading contribution in the strong-field limit comes from the choice $m_1 = m_2 = m_3 = m_4 = j$, where $\mathscr{J}^{}_0(z \rightarrow 0) \rightarrow 1 - z^2/4 + ...$. The correction to the leading contribution comes from the choice $j-m_1 = \pm 1, m_2 = m_3 = m_4 = j$, and the three equivalent choices obtained by exchanging the indexes $m_1, m_2, m_3, m_4$. In summary, in the strong-field limit, it is permissible to expand the range function $f(m_1,m_2,m_3,m_4)$ as follows,
\begin{align}
    f(m_1,m_2,m_3,m_4) &\approx \delta_{j,m_1} \delta_{j,m_2} \delta_{j,m_3} \delta_{j,m_4}  \nonumber \\
    &+ \frac{x}{2} \big( \delta_{j,m_1 \pm 1} \delta_{j,m_2} \delta_{j,m_3} \delta_{j,m_4} + 
    m_1 \leftrightarrow m_2 + m_1 \leftrightarrow m_3 + m_1 \leftrightarrow m_4 \big) 
    \nonumber \\
    &+ O(x^2)\ .
\end{align}
Inserting this expansion into Eq.~(\ref{HA}), we obtain that in the strong-field limit,  the Hamiltonian (\ref{int}) may be rigorously approximated by
\begin{align}
    \hat{H} &\approx F \sum_m m \; \hat{b}_m^{\dagger} \hat{b}^{}_m +
    V \sum_j  \; \hat{b}_{m}^{\dagger} \hat{b}^{}_{m} \hat{b}_{m+1}^{\dagger} \hat{b}^{}_{m+1} + \frac{xV}{2} \sum_m \hat{b}_m^{\dagger} \hat{b}^{}_m 
    \big(\hat{b}_{m-2}^{\dagger} \hat{b}^{}_{m-1} + \hat{b}_{m+2}^{\dagger} \hat{b}^{}_{m+1} + h.c.\big)
    \ .
 \label{ch}   
\end{align}

\begin{figure}
\includegraphics[width=0.25\textwidth]{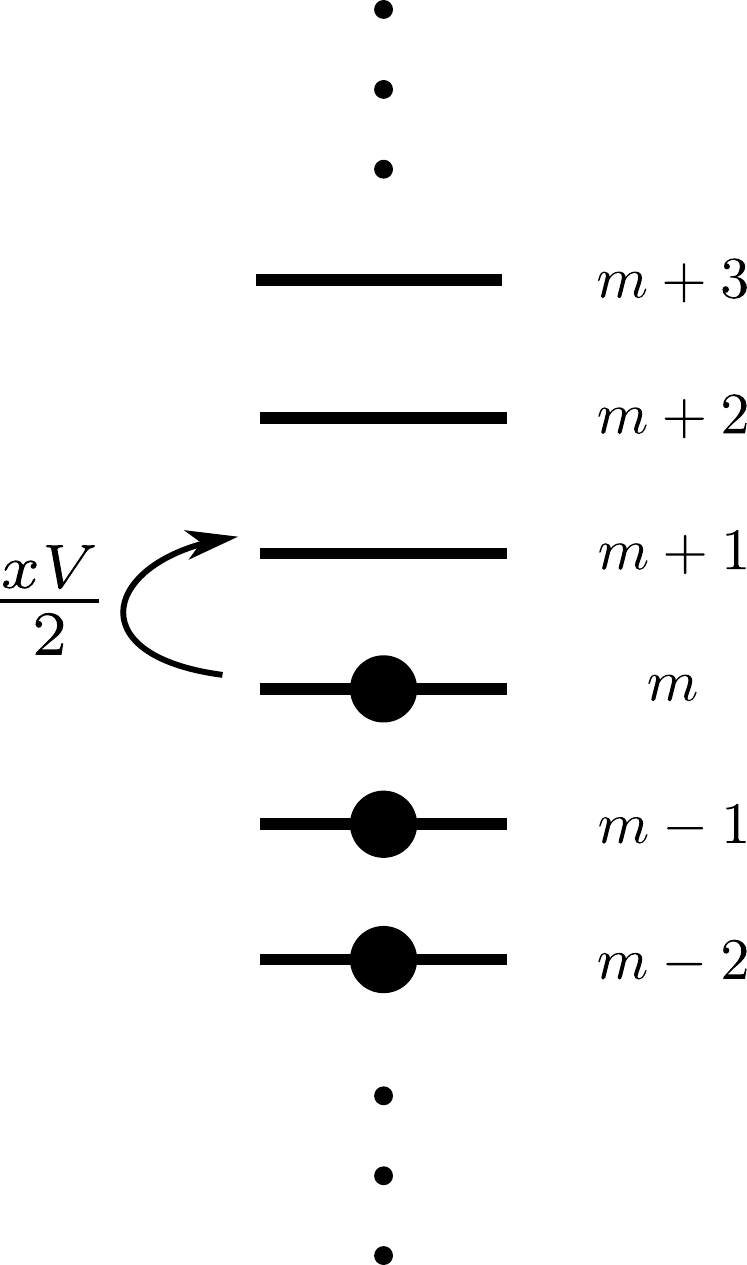}
\caption{Many-body hopping: If ladder-step $m-1$ is occupied, the particle at ladder-step $m$ may jump to ladder-step $m+1$.
The effective hopping strength is $xV/2 = V/F$.}
\label{MBH}
\end{figure}

\noindent which is similar to Eq. (21) in \cite{Krajewski}. The last term in Eq.~(\ref{ch}) provides a kind of \underline{many-body hopping} term (since the hopping only happens if site $m$ is occupied) that gives a hint of how the strong-field non-interacting ground-state with all sites till (above) a given site having unit (zero) occupancy is disturbed by the interaction $V$, moving electrons from the left towards the right side of the chain. This term will increase the characteristic length of the occupancy wall, as numerically observed. A schematic representation of this many-body hopping term is given in Fig.~(\ref{MBH}).



\twocolumngrid


\begin{thebibliography}{99}
\bibitem{Note} Reference \cite{NBY19} states (without derivation) that the localization length for the single-particle eigenstate $| m=0 \rangle$ is proportional to 
$[2 \sinh^{-1}(1/x)]^{-1}$, with $x=2/F$. This expression tends to $[\ln{(F/t)}]^{-1}$ in the large electric field limit,  and $t/F$
in the weak field limit.
\end{thebibliography}
\end{document}